\documentclass{iaus}
\usepackage{graphicx}
\newcommand\ep{\epsilon}
\newcommand\beq{\begin{equation}}
\newcommand\ee{\end{equation}}
\newcommand\apjl{\textit {ApJL}}
\newcommand\ts{\times}
\newcommand\apj{\textit {ApJ}}
\newcommand\pasp{\textit {PASP}}
\newcommand\mnras{\textit {MNRAS}}
\newcommand\araa{\textit {ARAA}}
\newcommand\nat{\textit {Nature}}
\newcommand\aj{\textit {AJ}}
\newcommand\aap{\textit {A\&A}}

\title[Magnetic Fields in Planetary Nebulae Paradigms and MHD Frontiers] 
{Magnetic fields in Planetary Nebulae Paradigms and 
Related MHD Frontiers}

\author[E.G. Blackman]   
{Eric G. Blackman$^1$}

{\affiliation{$^1$Department of Physics and Astronomy, University of Rochester,
Rochester, NY 14627, USA \break email: blackman@pas.rochester.edu}

\pubyear{2009}
\volume{259}  
\pagerange{100--100}
\date{"YOUR MAILING DATE"  and in revised form ??}
\setcounter{page}{119} \jname{Cosmic Magnetic Fields: From Planets,
to Stars and Galaxies} \editors{K.G. Strassmeier, A.G. Kosovichev \&
J.E. Beckman, eds.}

\begin{document}

\maketitle

\begin{abstract}
Many, if not all, post AGB stellar systems swiftly transition from a 
spherical to a powerful aspherical pre-planetary nebula (pPNE) outflow phase 
before waning into a PNe. The pPNe outflows require engine rotational energy and a mechanism to extract this energy into collimated outflows. 
Just radiation and rotation are insufficient but 
  a symbiosis between rotation, differential rotation  and large scale magnetic fields remains promising.  Present observational evidence for magnetic fields in evolved stars is 
suggestive of dynamically important magnetic fields, but both theory and
observation are rife with  research opportunity.
I discuss how magnetohydrodynamic outflows might arise in pPNe and PNe and   
distinguish different between approaches that address shaping vs. those that 
address both launch and shaping.
 Scenarios involving dynamos in single stars, binary driven dynamos, or accretion engines cannot be ruled out. One appealing paradigm involves accretion onto the primary post-AGB white dwarf core from a low mass companion whose decaying accretion supply rate owers first the pPNe and then the lower luminosity PNe.
Determining   observational signatures of  different MHD engines   
 is a work in progress.
 Accretion disk theory and large scale dynamos  pose many of their 
own fundamental challenges, some of  which I discuss in a broader context.

\end{abstract}

\firstsection 
\section{Introduction}

Asymmetries were observed in planetary nebulae PNe before the Hubble Space Telescope
but their ubiquity,  the rapidity
with which they develop, and collimated outflow power have
led to a cultural change in the field over the past decade (e.g. Balick \& Frank 2002).
Asymmetric  p/PNe have become the standard and 
understanding how these asymmetries arise is now 
fundamental rather than anecdotal. Perhaps most dramatic is  that 
collimated momenta of the pre-PNe (pPNe) (the reflection nebulae precursor to 
the ionized nebulae of PNe) exceed that which can be supplied by radiation
pressure alone (\cite{bujar01}). This is reminiscent of the  
young stellar object (YSOs) subject decades ago (\cite{yso}).  
The recent cultural change toward the view that the end states of stars 
(including supernovae e.g. Wang \& Wheeler 2008)
 are asymmetric, offers a plethora of research opportunity.

The beginning and ending of stars 
share the ingredients of  in-fall, collapse, turbulence, 
and angular momentum transport.
The associated increased free energy in differential rotation
  can be tapped to amplify fields and produce outflows, so
common underlying MHD principles  likely at work.
Focusing on pPNe/PNe, I review the role of
magnetic fields in these systems, and current evidence for their influence. 
I also address  broader  fundamental questions in magnetic field
generation and angular momentum transport for which pPNe are yet
another laboratory.  The role of large scale fields
and the need to connect 
different  MHD engines with observations
are central to the theme. 

\section{Basic Properties of p/PNe}

Generally, pPNe  exhibit a  fast bipolar outflow embedded within a 
slow spherically symmetric wind from the AGB star \cite{bujar01}.
Presently, data do not  rule out  all pPNe having gone through
a strongly asymmetric outflow and all PNe having gone through an 
asymmetric pNE phase.
Though AGB stars produce spherically symmetric outflows, the initial pPNe asymmetry
emerges within $\le 100$ yr (e.g. Imai et al. 2005).
PNe could reflect the late stages of the mechanism 
that produces pPNe with less asymmetry at very late times as  supersonic motions damp. 

For pPNe ( \cite{bujar01}), each fast  wind has a typical age 
$\Delta t\sim 10^2-10^3$yr,  speed $\sim 50$km/s,
 mass $M_f\sim 0.5M_\odot$, outflow rate, ${\dot M}_f\sim 5 \times 10^{-4}M_{\odot}/{\rm yr}$, 
 momentum $\Pi\sim 5 \ts 10^{39}{\rm g.cm/s}$, and mechanical 
luminosity $L_{m,f}\ge 8\ts 10^{35}{\rm erg/s}$ (can be as high as 
$10^{37}$erg/s).
The slow pPNe wind has an age  $\Delta t\sim 6\ts 10^3$yr, a speed $v_w\sim 20$km/s 
a mass   $M_s\sim 0.5M_\odot$, outflow rate, ${\dot M}_s\sim 10^{-4}M_{\odot}/{\rm yr}$, 
 momentum $\Pi_s \sim 2 \ts 10^{39}{\rm g\ cm/s}$, and mechanical 
luminosity $L_{m,s}\sim  10^{34}{\rm erg/s}$.
For PNe, observations suggest (Balick \& Frank 2002)
an  age 
$\Delta t\sim 10^4$yr 
a slow wind  of speed $v_s\sim 30$km/s 
of mass $M_s\sim 0.1M_\odot$, outflow rate, ${\dot M}_s\sim 10^{-5}M_{\odot}/{\rm yr}$, 
 momentum $\Pi_s \sim 6 \ts 10^{38}{\rm g\ cm/s}$, and mechanical 
luminosity $L_{m,s}\sim 3 \ts 10^{33}{\rm erg/s}$.v
PNe have  fast  winds of speed as high as $v_f\sim 2000$km/s,
 mass $M_f\sim 10^{-4}M_\odot$, outflow rate, ${\dot M}_f\sim 10^{-8}M_{\odot}/{\rm yr}$, 
 momentum $\Pi_f\sim 4 \ts 10^{37}{\rm g\ cm/s}$, and mechanical 
luminosity $L_{m,f}\sim 1.3 \ts 10^{34}{\rm erg/s}$.  

The pPNe phase  demands the  most power and  
the linear momenta of fast bipolar pPNe
outflows seems  too large for radiation driving \cite{bujar01}. 
This 
motivates the need to tap rotational energy, either from 
redistribution of angular momentum within a single star or 
via deposition of angular rotational energy from binaries. 


\section{Observations of  Magnetic Fields}

Detection of magnetic fields
in p/PNe is not by itself a proof of their dynamical importance as 
the relative strength of the  field and the local kinetic energies must be considered.
Complementarily, a weak magnetic field at large distances from the outflow engine  does not rule out a magnetically dominated launch at the engine. 

Magnetic field detection  is technically challenging 
but careful analysis has led to estimate of magnetic field strengths and geometries
in several important classes of late AGB and post-AGB systems by different techniques. 
In water maser sources such as W43 (an evolved AGB star with a precessing jet signaturing the beginning of the pPNe stage) magnetic fields 
of $\sim 85$mG have been inferred at radii of 500AU from the engine
via interpretation of circular polarization of $H_2O$ masers (Vlemmings et al. 2006). 
The 85mG fields at 500AU and their inferred geometry seem consistent with what is needed to  dynamically influence the flow. (See  Sabin et al. 2007 for field geometry in 
other sources).
 The precession suggests a binary interaction, but
whether the field is dynamo produced and/or 
connected to a companion within the envelope or the
AGB core is uncertain. 
 Vlemmings (2007) reviews maser field measurements of giant stars 
highlighting that the $H_2O$ masers 
probe  field on $\sim 10^2$AU scales, 
 OH masers probe $10^3$ AU scales and  SiO masers probe $\le 10$AU. 
The complied measurements show that, statistically, fields fall off faster
than $r^{-1}$ but slower than $r^{-3}$. This is  a weak constraint on models.

From VLT spectropolarimeteric analysis of several
central stars of PNe (CSPN), Jordan et al. (2005) detected
magnetic field strengths of order kG. The scale of the photosphere for
these system is $\sim 2\times 10^{10}$cm.
These field strengths are  are consistent with the larger scale scalings
presented in Vlemmings et al. (2007) and  the strengths of fields needed to
power the pPNe outflows by Poynting flux, and fields 
produced by dynamos in the AGB engines (Nordhaus et al. 2007).

Soker and Zoabi (2002) and Soker (2006b) have argued for weaker fields and rather than a primary driver of outflows, an indirect shaper of outflows possibly 
via influence on the dust distribution and geometry of radiative driven winds.
But this would leave unsolved the  large 
collimated momenta of pPNe engines {\cite{bujar01}) 
for which radiation driving is insufficient.
\section{Dynamical Magnetic Shaping and/or Launching} 
\label{sec:1}

The ``launch'' 
region 
of MHD  outflows is the region where the magnetic force  dominates the flow.  
This extends to a height typically no greater than  
$z_c\sim 50 R_{i}$, where $R_{i}$ is the inner-most radial scale of the engine.
(e.g. the inner radius of an accretion disk).
In the launch region the bulk flow is sub-Alf\'enic below  $z_c$.
The ``propagation'' region describes $z>z_c$
where the poloidal flow speed exceeds the Alf\'en speed, 
eventually approaching its asymptotic speed.
Presently, only the propagation regions are 
observationally spatially resolved. Also, 
because there are $\sim 2$ or more orders of magnitude between the
 engine where dynamos and jets formation operates and the asymptotic propagation region the computational demands prohibit simulatating the  combined  physics
of the launch region, jet formation and asymptotic propagation.
As emphasized in sec 7., even the nonlinear physics of just an accretion disk
has been prohibitively computationally expensive.
We must patch together different pieces of physics from different approaches
and scales  to extract a complete picture.

\subsection{Shaping with Imposed Fields}



In a  semi-analytic approach, Chevalier and Luo (1994),
imposed a magnetic field geometry such that the toroidal field
falls off more slowly than the poloidal field and at large distances and 
a spherical wind shocks the ambient gas increasing the toroidal field via compression.
For their parameters, the magnetic shaping  occurs only
after the shock.

Garc{\'{\i}}a-Segura
 (1997) and Garc{\'{\i}}a-Segura et al. (1999) 
were the first to simulate the direct effect of imposed toroidal fields
on PNe shaping by driving a hydrodynamic wind into 
a pre-magnetized  medium.
The  field strengths were consistent with those of sec. 3 above.
and their influence  (along with stellar rotation, and photoionization) on the shaping was  summarized in  simple schematic  (Garc{\'{\i}}a-Segura 1999) .
Rotation + magnetic fields leads to strong collimation compared to 
the weaker collimation of just the rotational influence on radiatively driven winds. 
Gardiner \& Frank (2001) discussed that the shaping observed in such
simulations were outside the restrictive parameter regime of
Chevalier \& Luo (1994) and that the influence of reasonable strength
 fields should generically affect the outflow  before the shock compression.
.


\subsection{Launching and Shaping with Imposed Initial Fields}

Because the observations require both a mechanism of launch and 
shaping, it is necessary to consider the role of magnetic fields in both.
 Garc{\'{\i}}a-Segura et al. (2005) 
 imposed a toroidal field  the surface of the AGB star to
launch an outflow via the field pressure gradient  and studied the propagation and 
asymptotic  collimation over
2 orders of magnitude in scale.  Given field strengths of $\sim 40G$ 
(consistent with observations) at the surface of the AGB star, outflows consistent with the required power and shaping were observed.  

Matt et al. (2006) added additional physics in 
 demonstrating how an initially weak field can grow linearly in time 
by extracting
rotational energy of its anchoring base,  and produce a rapid 
bipolar outflow from the gradient in toroidal field pressure.
They took a gravitating spherical core
surrounded by an initially hydrostatic envelope of ionized gas. A dipolar magnetic 
ﬁfield was anchored on the core, threading the envelope. The  
core was set to rotate initially at 10\% of the escape speed.
The toroidal field amplified from the differential rotation between core and 
envelope and when the toroidal
field pressure gradient overcame the envelope binding energy,
envelope material was rapidly expelled in a quadrupolar outflow:
The wound up diploe field had maximum toroidal field 
 at intermediate poloidal angles in each hemisphere, so material was 
both   along the poles  also squeezed out
from the equator. 

\subsection{Launching via Dynamo Engines} 

The models of the previous two subsections do
not fully address where the dynamically important large scale
fields come from. Even the Matt et al. (2006) paper invoked
linear, laminar field growth and treating the rotator base as a boundary condition.

Papers such as those by Pascoli (1993,1997) Blackman et al. (2001a,b), 
Nordhaus et al. (2007) use
flavors of a  mean field dynamo theory for stellar 
or disk engines to estimate field strengths and Poynting fluxes that could
arise.  These papers are not simulations and do not track global field
geometry, or follow the production and evolution of outflows from the field production region. Furthermore, the nonlinear physics of dynamos (see sec. 7) 
 require approximations when emedding them in 
astrophysical scenarios.
Despite their shortcomings, such toy models do produce promising results
with respect to field strengths and Poynting fluxes

\section{Single Star  vs. Binary Models}

\subsection{Single star models not yet ruled out}

Binaries can easily supply the needed free energy to amplify fields via differential rotation
(ordhaus \& Blackman 2006). 
Isolated star MHD models cannot be ruled out, but there are caveats
 (Soker 2006b; Nordhaus et al. 2007) as I now discuss.

Blackman et al. (2001) investigated an isolated interface dynamo model operating at the base of the AGB convective zone in which angular momentum is conserved on spherical shells as the star evolves off the main sequence and the resulting rotation profile provides the available differential rotational energy from which the field is amplified. To drive bipolar pPN/PN, the corresponding dynamo must operate through the lifetime of the AGB phase ($10^5$yr) until radiation pressure has bled most of the envelope material away. Only then can the Poynting flux unbind the remaining material. 
But the dynamo would drain differential rotation
on time-scales short compared to the AGB lifetime. 
Only if differential rotation is extracted from deep within the core or re-seeded by convection can the drain be overcome (Nordhaus et al. 2007). 
 Viability of the single star dynamo outflow model, depends on  whether or not the differential rotation can be re-supplied over the AGB lifetime via convection as in the Sun. 

\subsection{Common envelope binary scenarios}

In a common envelope (CE) \cite{il93} the companion drags on
the envelope of the primary, transferring angular momentum and energy.
If the envelope cooling time is long enough, a
 fraction $\alpha \ge 0.1$ of the loss in gravitational energy
 can spin up and unbind the envelope.
\cite{RRL} discuss  the   secondaries  
for which accretion disks will form around the primary core. 
Brown dwarf (BD) ($0.003M_\odot <M_{crit}\sim 0.07M_\odot$) 
radii increase with decreasing mass while their Roche
radii decrease with decreasing mass. 
Such objects unstably lose mass. Since the
circularization radius lies outside of the primary's core, 
a disk can form within a few orbits. This contrasts the $M>M_{crit}$ case for 
which the stellar radius
decreases with decreasing mass more strongly than the Roche
radius. Supercritical companions 
have a circularization radius {\it within} the
primary's core. Material leaving the secondary would then 
initially spiral swiftly into the primary rather than orbit quasi-stably.
Nordhaus \& Blackman (2006) show  that accretion from planets
and low mass stars may also be important 
(and perhaps more common 
because of the brown dwarf desert \cite{gl06})
and can supply the needed accretion power for pPNe.
Planets are of sub-critical mass and a disk could form as per a BD,
before unbinding the envelope.  

A companion star is supercritical so if stellar incurs Roche lobe overflow after CE,
sustained accretion requires the binary to lose
angular momentum.
This could  happen 
as the  overflowing secondary drags on  residual inner envelope material.
Even though the circularization radius for $M_2 > M_{crit}$ 
is inside the core, the in-spiraling material
still incurs differential rotation and could amplify magnetic fields.
A significant energy release via accretion 
in the $M_2 > M_{crit}$ case could occur on impact to 
the inner core, producing dwarf novae bursts
masked by the stellar envelope. 
If this initial accretion phases can drop  $M_2$   below $M_{crit}$, 
keep $M_2$ filling its Roche lobe, and  leave enough angular
momentum to form a Keplerian disk, then  
accretion  could   proceed  as for the initial $M_2<M_{crit}$ 
case.

\cite{huggins} suggests a $\sim O(100)$ year delay  between 
ejection of  cicumbinary dust tori and jets in pPNe.
If  a fraction of ejected envelope material
becomes the dust torus, the delay 
could be the time for the companion to lose
enough mass to move the  circularization radius
outside the core for the supercritical case, or a viscous time.

\section{Accretion Disk Outflows in pPNe and PNe:}
\label{sec:5}

Accretion disk outflows have a mechanical power 
 \cite{bfw01}
\beq
L_{m}\sim {GM_*{\dot M}_{a}\ep \over 2 R_i} = 4.5 \ts 10^{36} \ep_{-1}{M_*{\dot M}_{-4}\over R_{i,10}},
\label{1}
\ee
where $\ep$  is the
efficiency of conversion from accretion to outflow,
$R_i$  is the inner disk radius, ${\dot M}_a$  is the accretion rate,
$G$ is Newton's constant, $M_*$ is the central stellar mass,
 $R_{i,10}\equiv R_i/10^{10}{\rm cm}$, 
$\ep_{-1}\equiv \ep/0.1$ and ${\dot M}_{a,-4}= {\dot M}_a/10^{-4}M_\odot$/yr.
For an MHD outflow, Eq. (\ref{1}) equals the Poynting flux at the launch surface.
Propagation into a low density medium
produces an asymptotic outflow speed  $\sim \Omega r_A$  \cite{pudritz04}
where $\Omega$ is the angular speed of field anchor point  and $r_A$ is the
radius where the  poloidal outflow speed 
equals the Alfv\'en speed. This product is typically a few times the escape speed of the inner most radius of a disk and is thus at least 
$
v_{out}\sim v_{esc} =1600 \left(M_*\over R_{*,10}\right)^{1/2}{\rm km/s}$

\subsection{Accretion onto Primary}

The $v_{out}$ above
depends only weakly on ${\dot M}_a$ via $R_i$, but strongly
on the inertia of material 
blocking the outflow:  Momentum conservation gives 
\beq
v_{obs}={M_{f}v_{out}
\over f_\Omega M_{env} + M_{f}}
\sim 80{\rm km/s},
\label{3}
\ee
where 
$M_{f}/ M_\odot=3.3 \times 10^{-4}\ep_{-1}{\dot M}_{a0,-3}\int_1^{1000} \tau^{-5/4}  d\tau
$ is the mass in one of the fast collimated  outflows, $M_{env}$ is the envelope mass,  $f_\Omega\sim 0.2$ is the solid
angle fraction intercepted by the collimated outflow, and
$\tau\equiv  t/1yr$ is used to incorporate
 ${\dot M}_a\propto t^{-5/4}$ of \cite{RRL}.
The numbers are scaled to pPNe 
 so   $M_{env}>> M_{f}$ and for an envelope of mass $2 M_\odot$,
the intercepted mass is 0.2 $M_\odot$ for $f_\Omega=0.2$.

Eq. (\ref{3}) is the observed speed of the fast 
when blocked and loaded by the 
envelope. By the end of the pPNe phase, the envelope is quite extended,  
reducing the optical depth and revealing material
moving at the ``free streaming'' fast wind speed. 
Assuming a dust-to-gas mass ratio of 1/100 and 
micron sized grains of density of 2g/cm$^3$, 
the optical depth from dust is 
$\tau_d \sim 2.5 \ts 10^{-3}\left({n_d\over 2.5 \ts 10^{-13}{\rm cm^{-3}}}\right)
\left({\sigma_d\over  10^{-8}{\rm cm^2}}\right)
\left({R\over  10^{18}{\rm cm}}\right),
$
scaled for PNe.
For pPNe, the density increases by  $\ge 10^4$
and $R$ is down by a factor of 10, so $\tau_d\ge 2.5$.
The different optical depths of pPNe and PNe  can thus explain
 why observed PNe fast winds can have 
$v_f=v_{out}>1600$km/s, whilst those of pPNe have $v_f=v_{out}<100$km/s. 

Time dependent  accretion outflows described with 
Eqs. (\ref{1}) and (\ref{3}) are consistent  with the high pPNe outflow mechanical luminosity  and the fast PNe wind speed  
of Sec. 1 when $M_*/R_*$ corresponds to a  WD.
 \cite{RRL}
considers  a  companion of mass $M_2\sim 0.03M_{\odot}<M_{crit}$ 
and a Shakura-Sunyaev {\cite{ss73} viscosity parameter $\alpha_{ss}\sim 0.01$, 
for which the accretion rate then decays as 
${\dot M}_a\sim 1.6\ts 10^{-3}  t^{-5/4}{M_\odot}/$yr..
Using this in (\ref{1}) with $\ep=0.1$ 
for $t=100$ yr with $R_i= 2\times 10^9$cm
and $M_1=0.6M_\odot$ gives $L_{m,f}\sim 4.3 \times 10^{39}(t/{\rm 1yr})^{-5/4}$.
This provides the needed power demands of Sec. 1  for pPNe  after $1000$ yr and
for PNe after $10^4$ yr.  

A more careful analysis of the predicted jet speed evolution is needed for 
specific outflow models as the envelope evolves in order to test the idea contained within the rough estimates just discussed. See also Garicia-Diaz et al. (2008) for a non-accretion based time dependent comparison of jet outflow speed from simulations with observations.





\subsection{Accretion onto Secondary}
It is also possible for accretion disks to form around the secondary
(Soker \& Livio 1994; Mastrodemos \& Morris 1998; Soker 2005).
The Bondi wind accretion rate is 
${{\dot M}\over {\dot M}_s}=\left(M_2\over M_1\right)^2{(v/v_s)^4\over 
[1+(v/v_s)^2]^{3/2}},
$ where $v$ is the orbital speed of the secondary and
$v_w$ is the slow wind speed from the primary.
In general, for $M_2< M_1$, reasonable parameters
provide an accretion rate compatible 
PNe luminosities  if the companion is either main sequence or compact
star, but the ubiquity of high collimated fast wind pPNe powers (\cite{bujar01})
would require an overabundance of accreting WD companions.

\section{Key Issues in Large Scale MHD Dynamo and Accretion Theory}
\label{sec:dynac}


\subsection{Large Scale Dynamo Theory: Recent Developments, Open Questions}

Large scale dynamo (LSD) theory describes the sustenance of magnetic fields  on time/spatial scales larger than  turbulent scales.
Whether large scale fields for jets are  advected or LSD produced  has been debated but  but ultimately, the equations that include a 
the competition between  turbulent transport, flux accretion, as well as LSD action
 need to be solved. Field reversals in the sun prove that an LSD can and must operate therein.

For $\sim$ 50 years, a problem with textbook LSD theory (e.g. \cite {moffatt})
has been the absence of a proper saturation theory 
that predicts how strong the large scale fields get
before non-linearly quenching  via the backreaction of the field on the driving flow kicks in. But substantial progress toward a nonlinear 
 mean field theory has emerged  in the last decade via a symbiosis between 
analytical and numerical work. 
Coupling the dynamical evolution of magnetic helicity into the dynamo equations 
turns out to be fundamental for predicting the saturation seen in simulations.
For recent extensive reviews see
(Brandenburg \& Subramanian 2005; Blackman 2007).

Much of the  work in LSD theory, has focused on systems that are initially globally reflection asymmetric (GRA). This means a global pseudoscalar is 
imposed by the boundary conditions--for example, rotation and stratification lead to the kinetic helicity 
pseudoscalar, common to the standard textbook (\cite{moffatt})
``$\alpha_{dyn}$ effect'' of 
 mean field dynamos.  However magnetic helicity is 
actually the unifying quantity for LSD.
There are two  classes of GRA LSD for which
an electromotive force aligned with the mean magnetic field is essential (see  e.g. Blackman 2007).
The first is flow driven helical dynamos (FDHD) which occur inside of astrophysical rotators. Here the initial energy is dominated by flows and the field responds.  
These are linked to a corona by magnetic buoyancy. In coronae, the second type of LSD, the magnetically driven helical dynamo (MDHD) can operate.  This characterizes relaxation of magnetic structures to larger (jet mediating) scales in a magnetically dominated environment subject to the injection of smaller scale magnetic helicity.
The MDHD is directly analogous to laboratory plasma
dynamos that occur in reverse field pinches (RFPs) and Spheromaks.

LSDs always involve some helical growth of the large scale field which is coupled
to a helical  scale fields of opposite sign. 
When small scale magnetic or currently helicity evolution is coupled to the large scale 
field growth, the modern mean field `dynamo $\alpha_{dyn}$ effect that
predicts the correct satuation becomes the difference between kinetic helicity and 
current helicity:  For a FDHD simulated in a closed box (Brandenburg 2001), 
the current helicity builds up as the large scale field 
grows and quenches the FDHD (Blackman \& Field 2002). 
Complementarily, for an MDHD, the system is first dominated by the 
current helicity and a growing kinetic helicity then acts as the backreaction 
(Blackman \& Field 2004). 
Both
FDHD and MDHD are accessible with in the same framework, all unified by tracking magnetic helicity evolution, and aided by thinking of the field as ribbons rather than lines
(Blackman \& Brandenburg 2003).
More work on how the fields evolve from within
the rotator to produce the global scale fields in  coronae which in turn produce
jets are needed as most work on dynamo theory has focused on the FDHD.

Because  the buildup of small scale magnetic (or current) helicity 
quenches the LSD, preferential ejection of 
 small scale helicity vs. large scale helicity through a boundary
can in principle alleviate this quenching 
(Blackman \& Field 2000; Vishniac Cho 2001; Sur et al. 2007).  
Numerical simulations  support this general notion, particularly when shear is present 
and when  surfaces of shear align toward open boundaries
(Brandenburg \& Sandin 2004; Kapyla et al. 2008), thereby allowing  needed helicity
fluxes.

LSD action has also been observed in non GRA simulations 
(e.g. Yousef et al 08; Lesur \& Ogilvie 08)
implying that the minimum global ingredients for this class of analytic LSD
is shear, plus turbulence that feeds azimuthal field back to toroidal field.
There is work in progress to understand 
the simulations guided by analytic models (e.g. Vishniac \& Brandenburg 1997; Blackman 1998;
Kleeorin \& Rogachevskii 2003;  Schekochihin et al 2008).
The non GRA LSDs grow large scale fields on scales larger than the turbulence but smaller than the global scale. In coherence regions, there is 
a field aligned electromotive force (EMF), and thus  an intermediate scale source
of magnetic helicity that may switch signs between coherence regions and globally average to zero.  It may be that the non-GRA LSD action always involves a local helicity flux between coherence regions.

\subsection{Accretion Disks: More questions and the need for large scale fields}

The magneto-rotational instability (MRI) has emerged 
as a leading candidate for angular momentum transport in 
 accretion disks
(e.g. Balbus \& Hawley 1998) 
Although the MRI exists without the LSD, they are likely coupled in nature.
I explain  this below.

There is a disconnect between what 
shearing box simulations have told us about the 
MRI vs. how the instability might operate in nature.  To date, simulations have primarily told us  that the MRI is plausible but do not produce a robust theory of saturation or 
robust values of transport coefficients for modelers. 
This may frustrate, but patience (for several more decades) is required, 
as the computational and conceptual demands are substantial.
For example, to achieve better resolution, 
most first generation MRI simulations (except Brandenburg et al.1995)
did not use explicit viscosity or magnetic diffusivity
(see discussion in  e.g. Fromang et al. 2007).
the magnetic Prandtl number affects the magnetic spectrum
and the transport coefficients. In adddition, 
the value  of the  angular momentum transport coefficient $\alpha_{ss}$ depends strongly on the box size and the strength of the initially imposed weak mean field strength (Pessah et al. 2007). Interestingly, $\alpha_{ss}$ varies $\sim 4$ orders between simulations, 
but $\alpha_{ss}\beta$, where $\beta$ is the ratio of thermal to magnetic pressure, 
is nearly a constant (Blackman et al. 2007).

 In addition to the need for explicit diffusivities, perhaps
the most important frontier is actually
role of  large scale magnetic fields in angular momentum transport
and thus, non-local contributions to magnetic stresses that transport angular momentum.
Its importance is motivated  from three different paths: 

(1) Large scale fields
are evidenced from theory, simulation, and observations of  
coronae and jets. Large field structures more easily
survive the buoyant rise to coronae without being shredded by turbulence
within the disk. Plausibly, the integrated stress for structures on scales above that which survives 
the buoyant rise would contribute to the large scale magnetic stress.
But shearing boxes that study only a local region 
non-local large scale magnetic stress is excluded. In a real system this
could be the most important part.

(2)  Shearing box simulations artificially 
impose a steady-state because differential rotation is imposed as 
a steady forcing not subject to the backreaction 
of the amplified field. Hubbard and Blackman (2008b),
 argue that this may be more restrictive that previously recognized: 
In a real disk, energy in differential rotation is susatined  only by accretion 
itself. If a steady-state is to be maintained via turbulent transport alone
then there must be 100\%  
power throughput from differential rotation to the turbulent cascade.
This is not guaranteed if large scale fields drain power.  An accessible  
steady-state solution must then incorporate stresses from large-scale fields in addition
to turbulent transport.

(3) The role of large scale magnetic fields is consistent with the implications of  
(Pessah et al. 2007) which shows that 
MRI stresses in simulation boxes where the radial extent is of order the vertical scale height 
scale with the ratio of box size to scale height. 
 The contribution of large scale fields would  increase as the radial and vertical scales are increased,  highlighting a strongly non-local contribution to stresses that transport angular momentum.  Evidence for non-local MRI behavior is also seen in \cite{bodo2008}.  
Box sizes for thin disks must be extended in the radial direction and
accordingly in the vertical direction as buoyant loops tend to have radial
scales comparable to vertical scales. 


Ultimately, mean field accretion disk theory should be coupled to an
LSD theory in a real disk as they are actually artificially separated 
components of what should be a single mean field theory. Note that 
for any turbulent disk, any  assumption of axisymmetry for bulk quantities
automatically requires that theory to be a mean field theory. 
 This is  often  veiled by a mere replacement of the actual viscosity with a turbulent viscosity as in the  Shakura-Sunyaev approach (\cite{ss73}).
 Hubbard \& Blackman (2008a) for example, suggest that in 
a formal mean field theory,
a term involving fluctuations of density and
 velocity might be interpreted as an additional transport coefficient
in the mean surface density equation, restricting available steady states. 
Balbus et al. (1994)incorporated this term into a redefinition of accretion rate.

\section{Much work needed to link theory and observation}
\label{sec:5}

It remains  a major 
 challenge to rigorously couple the engine physics of field generation and
accretion to jet launch and jet propagation in a unified theory or simulation
 that make distinct observational predictions for specific engines.
This enterprise spans several subfields of theoretical astrophysics,
 let alone the specific application to p/PNe.  

Here however, is a list of possible lower hanging fruit for linking theory with observations:(1) Evaluate the kinematic constraints/predictions of outflows from
the scenario of accretion onto the primary and compare the distribution of inferred fast 
outflow speeds to what would be expected from 
known binary statistics of low mass stars. 
This would help constrain the commonality of  accretion onto the primary vs. secondary 
as the latter has a broader range of masses.
(2) The more massive the companion, the more
CE models would  predict mostly  Oxygen rich rather than Carbon rich 
post AGB systems because the binding energy for the early AGB phases is higher.
Low mass companions like planets 
may terminate the AGB only in the thermal pulse phase. 
(3) Crystalline dust  in post-AGB systems can be  produced if a binary
induced spiral shock anneals silicates \cite{edgar}. Is this universal?
(4) CE evolution would predict equatorial outflows
that precedes any accretion driven
poloidal jet. Is this consistent with 
the delay of \cite{huggins} and the geometry of equatorial outflows?
(6) Are fast outflows contaminated by material that could represent 
accretion disk residue of  shredded low mass companions?
(7) Are time scales of observed outflow 
precession consistent with the gravitational
influence of a binary?
(8) Can  double peaked line profiles be detected
to identify accretion disks within the launch region?
(9) Can  shrouded novae outbursts from a $M_2>M_{crit}$ companion feeding the primary be detected in X-rays?
(10) Improved statistics on the fraction of
 bipolar pPNe, the fraction of suitable precursor binaries for CE,
and the fraction of stars which evolve to be pPNe 
will improve evaluation as to whether  all PNe incur asymmetric phase.
(12) Can the approach of Ferreira et al. (2006)  be generalized to pPNe  outflows?


\begin{discussion}
\end{discussion}

\end{document}